\documentclass[11pt,a4paper]{article}
\usepackage[margin=25mm]{geometry}
\usepackage[T1]{fontenc}
\usepackage{lmodern}
\usepackage{authblk}

\usepackage{amsmath,amssymb,amsthm,braket}

\usepackage{enumitem}
\setlist{itemsep=3pt,topsep=5pt}

\usepackage[dvipsnames]{xcolor}
\usepackage{tikz}
\usetikzlibrary{arrows.meta,decorations.pathmorphing}
\usepackage{subcaption}
\usepackage{booktabs}

\definecolor{ptAmber}{HTML}{B87500}%
\definecolor{ptBlue}{HTML}{2367A0}%
\definecolor{ptInk}{HTML}{202A34}%
\definecolor{ptMuted}{HTML}{596570}%

\usepackage{url,hyperref}
\usepackage[noabbrev,nameinlink]{cleveref}

\hypersetup{
  colorlinks=true,
  allcolors=MidnightBlue
}

\newcommand{\complexityclass}[1]{\ensuremath{\mathsf{#1}}}

\NewDocumentCommand{\QMA}{o}{%
  \ensuremath{\complexityclass{QMA}\IfValueT{#1}{(#1)}}}

\NewDocumentCommand{\dQMA}{o}{%
  \ensuremath{\complexityclass{dQMA}\IfValueT{#1}{(#1)}}}

\NewDocumentCommand{\dQMAsep}{o}{%
  \ensuremath{\complexityclass{dQMAS}\IfValueT{#1}{(#1)}}}

\newtheorem{lemma}{Lemma}[section]
\newtheorem{theorem}[lemma]{Theorem}
\newtheorem{corollary}[lemma]{Corollary}
\theoremstyle{plain}
\newtheorem{result}{Result}
\crefname{result}{result}{results}
\Crefname{result}{Result}{Results}

\theoremstyle{definition}
\newtheorem{definition}{Definition}[section]
\theoremstyle{plain}

\newtheorem*{repeatedlemma}{\repeatedlemmaname}

\newcounter{protocol}
\newsavebox{\protocolbox}

\newenvironment{protocol}[1]{%
  \par\medskip\refstepcounter{protocol}%
  \setlength{\fboxsep}{8pt}%
  \begin{lrbox}{\protocolbox}%
  \begin{minipage}{\dimexpr\linewidth-2\fboxsep-2\fboxrule\relax}%
  \begin{center}
    \bfseries Protocol \theprotocol. #1
  \end{center}%
}{%
  \end{minipage}\end{lrbox}%
  \noindent\fbox{\usebox{\protocolbox}}\par\medskip
}

\crefname{protocol}{protocol}{protocols}
\Crefname{protocol}{Protocol}{Protocols}

\usepackage{authblk}

\author[1]{Alex Della Schiava}
\author[1]{Atsuya Hasegawa}
\author[1]{François Le Gall}
\author[2]{Harumichi Nishimura}

\affil[1]{Graduate School of Mathematics, Nagoya University, Japan}
\affil[2]{Graduate School of Informatics, Nagoya University, Japan}
\title{Reducing Provers in Distributed Quantum Verification}

\date{}

\begin{document}

\maketitle
\begin{abstract}
In the distributed quantum Merlin--Arthur ($\dQMA$) model introduced by Fraigniaud, Le~Gall, Nishimura, and Paz (ITCS~2021), a network of quantum computers verifies a global property with the aid of a quantum proof supplied by an untrusted prover. 
While the nodes are computationally unbounded, the challenge is to limit proof size and the communication required for verification. In this paper, we consider the model of Hasegawa, Kundu, and Nishimura (PODC~2024), where the proof must be unentangled across nodes, and define its extension to $k$ unentangled provers, denoted by $\dQMAsep[k]$. In this setting, we investigate the proof size per node and the communication costs of reducing the number of provers.

Our first result reduces the number of provers from $k$ to two, doubling the proof size without additional
communication. Moreover, with four additional rounds, we can reduce the number of provers from \(k\) to one, using at most three times the original local proof size plus a logarithmic term in the number of nodes.
Finally, we establish amplification for perfectly complete protocols. In this regime, a single prover suffices with a constant-factor overhead in our reductions, yielding \(\dQMAsep=\dQMAsep[k]\).

\end{abstract}

\section{Introduction}\label{sec:introduction}
Distributed verification asks how nodes can verify a global property of a network or its inputs despite having only local information.
In proof-labeling schemes, introduced by Korman, Kutten, and Peleg~\cite{KKP10}, and locally checkable proofs, studied by Göös and Suomela~\cite{GS16}, an untrusted prover assists the nodes by supplying a proof, which they check through local computation and communication. While the nodes are computationally unbounded, the challenge is to limit proof size and the communication required for verification.

Fraigniaud, Le Gall, Nishimura, and Paz~\cite{FGNP21} introduced a quantum counterpart of this setting: \emph{distributed quantum Merlin--Arthur} protocols, or $\dQMA$.
The proof is now a quantum state, and verification proceeds in synchronous rounds. In each round, the nodes perform local quantum operations and exchange quantum messages with their neighbors. The protocol accepts if and only if all nodes accept. On a yes-instance, some witness is accepted with high probability; on a no-instance, every witness must lead to rejection by at least one node with high probability.

The quantum setting can substantially reduce the resources required for distributed verification: for equality on fixed-size networks, it offers exponentially smaller proofs~\cite{FGNP21}.
Le~Gall, Miyamoto, and Nishimura~\cite{GMN23} developed distributed state synthesis and showed that additional resources can compensate for restricting communication to classical messages.
They also reduced constant-turn classical interactive proofs to five quantum turns with constant-factor overhead in proof size~\cite{GMN23interactive}.
Hasegawa, Kundu, and Nishimura~\cite{HKN24} established further protocols and lower bounds, and proposed studying a variant of $\dQMA$ in which $k$ provers supply mutually unentangled proofs, denoted $\dQMA[k]$.

In the centralized setting, Kobayashi, Matsumoto, and Yamakami~\cite{KMY03} introduced $\QMA[k]$, where a polynomial-time quantum verifier receives mutually unentangled quantum witnesses from $k$ provers. Ignoring all but one proof gives $\QMA\subseteq\QMA[k]$, but whether additional provers offer greater power is a longstanding open question~\cite{Gha24, JLW26}. In the classical setting, the answer is negative: $k$ witnesses can always be concatenated into one polynomial-length proof and checked as before.
In contrast, combining \(k\) quantum witnesses into one proof allows entanglement between the registers, introducing cheating strategies for which the original soundness guarantee need not hold. Harrow and Montanaro~\cite{HM13} showed that two provers always suffice: $\QMA[2]=\QMA[k]$ for every polynomially bounded $k\ge2$. Whether a single prover suffices, that is,
$\QMA[2]=\QMA$, remains open.

In the centralized setting, the main question is whether prover reduction preserves polynomial-time verification, allowing polynomial overheads in proof size and computation.
In distributed verification, finer resource bounds matter: reducing the number of provers may increase global and local proof sizes or require additional communication. We therefore ask:
\begin{center}
\emph{What is the cost of reducing the number of provers
in distributed quantum verification?}
\end{center}

\subsection{Our model}\label{sec:our-models}

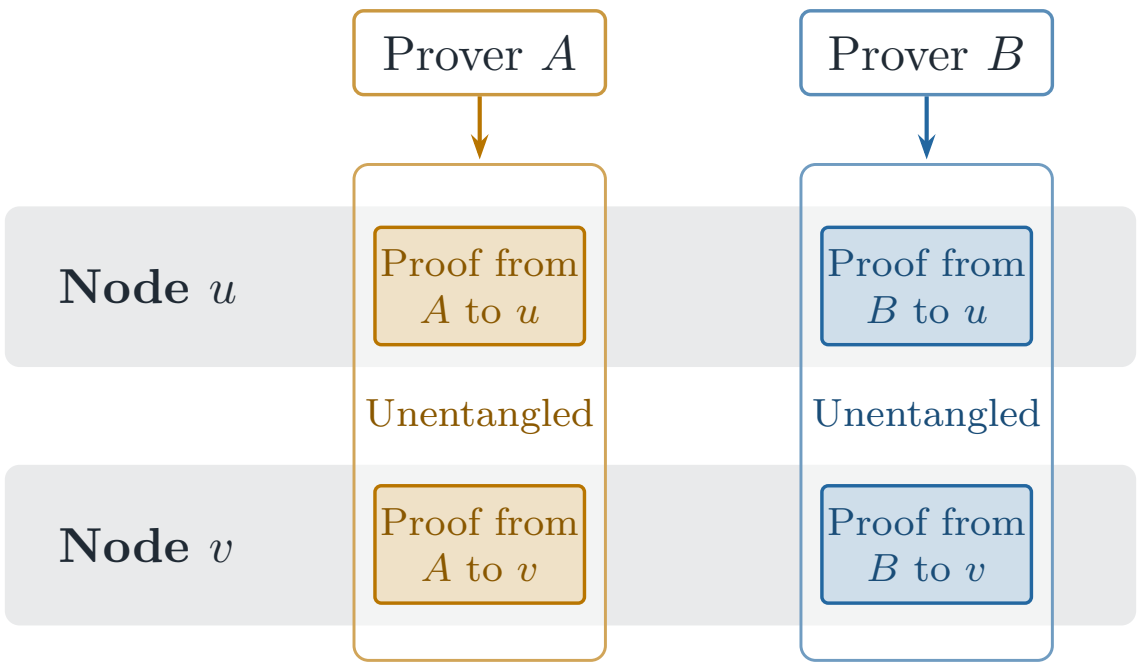
\begin{figure}
    \centering
\usetikzlibrary{arrows.meta}
\begingroup%
\definecolor{routingAmber}{HTML}{B87500}%
\definecolor{routingBlue}{HTML}{2367A0}%
\definecolor{routingInk}{HTML}{202A34}%
\resizebox{.485\linewidth}{!}{%
\begin{tikzpicture}[
  x=1cm,y=1cm,
  model reg/.style={draw=#1,fill=#1!22,line width=.7pt,
    rounded corners=1.3pt,minimum width=12.5mm,minimum height=5.5mm,
    inner sep=0pt},
  model prover/.style={draw=#1!75,fill=white,line width=.7pt,
    rounded corners=2pt,minimum width=18mm,minimum height=6mm,
    inner sep=3pt,font=\small,text=routingInk,align=center},
  model verifier/.style={font=\small\bfseries,text=routingInk,align=left},
  model label/.style={font=\scriptsize,align=center,text=#1!75!black,
    inner sep=2pt},
  model route/.style={-{Stealth[length=1.8mm,width=1.25mm]},
    line width=.85pt,shorten >=1pt},
  model enclosure/.style={draw=#1!65,fill=#1!7,line width=.6pt,
    rounded corners=3pt}
]
  \path[use as bounding box] (0,.75) rectangle (8.4,5.42);
  \fill[routingInk!10,rounded corners=3pt]
    (.15,2.85) rectangle (8.25,4.0);
  \fill[routingInk!10,rounded corners=3pt]
    (.15,1.0) rectangle (8.25,2.15);
  \node[model verifier,anchor=west] at (.4,3.43) {Node $u$};
  \node[model verifier,anchor=west] at (.4,1.58) {Node $v$};

  \draw[model enclosure=routingAmber,fill=white,fill opacity=.5]
    (2.65,.75) rectangle (4.45,4.3);
  \draw[model enclosure=routingBlue,fill=white,fill opacity=.5]
    (5.85,.75) rectangle (7.65,4.3);
  \foreach \cy/\vertex in {3.43/u,1.58/v}{
  \draw[model reg=routingAmber]
    (2.8,\cy-.42) rectangle (4.3,\cy+.42);
  \node[model label=routingAmber] at (3.55,\cy)
    {Proof from\\$A$ to $\vertex$};

  \draw[model reg=routingBlue]
    (6.0,\cy-.42) rectangle (7.5,\cy+.42);
  \node[model label=routingBlue] at (6.75,\cy)
    {Proof from\\$B$ to $\vertex$};
}
  \node[model label=routingAmber] at (3.55,2.5) {Unentangled};
  \node[model label=routingBlue] at (6.75,2.5) {Unentangled};

  \node[model prover=routingAmber] (pa) at (3.55,5.10) {Prover \(A\)};
  \node[model prover=routingBlue] (pb) at (6.75,5.10) {Prover \(B\)};
  \draw[model route,draw=routingAmber] (pa.south) -- (3.55,4.3);
  \draw[model route,draw=routingBlue] (pb.south) -- (6.75,4.3);
\end{tikzpicture}%
}%
\par
\endgroup%
    \caption{The $\dQMAsep[2]$ model on a two-node network.
    Amber and blue represent the proofs sent by provers
    $A$ and $B$, respectively. The two proofs are
    unentangled, and each proof is separable across nodes. Gray bands span the registers held by nodes \(u\) and \(v\).}
    \label{fig:models}
\end{figure}

We consider \(\dQMAsep[k]\) protocols, where \(k\) unentangled
provers each supply a proof that is separable, i.e.,
unentangled across nodes. Our interest in this model is motivated by the practical challenges
of distributing entanglement across a network~\cite{WEH18, AEE+23}.
\Cref{fig:models} illustrates this setting for $k=2$ provers. For a single prover, this coincides with the $\dQMA^{\mathrm{sep},\mathrm{sep}}$ model of~\cite{HKN24}, which we denote by $\dQMAsep$. We use the notation
\[
    \dQMAsep[k][c,s;\,p,r],
\]
where $c$ and $s$ are the completeness and soundness parameters,
respectively. The parameter $p$ bounds the \emph{maximum local proof size},
counting all proof qubits supplied to any one node by all
provers combined, while $r$ bounds the number of synchronous
communication rounds used throughout the protocol verification.

\subsection{Our results}\label{sec:our-results}

In the $\dQMAsep[k]$ model, we quantify the proof and communication costs to reduce the number of provers in a given network of \(N\) nodes.

Our construction immediately yields a distributed analogue of the two-prover reduction of Harrow and Montanaro~\cite{HM13}. By doubling the local proof size, we  reduce \(k\) provers to two without any communication overhead.

\begin{result}[\Cref{lem:product-test}, Two-prover reduction]\label{res:sep-two}
    For every integer \(k\ge2\) and completeness--soundness parameters \(c,s\) satisfying \(c>(6+s)/7\),
    \[
     \dQMAsep[k][c,s;\,p,r]
     \subseteq
     \dQMAsep[2]\!\left[
     c,\frac{6+s}{7};\,2p,r
     \right].
    \]
\end{result}

With four additional rounds, we can reduce the number of provers from \(k\) to one, using at most three times the original local proof size plus a logarithmic term in the number of nodes.
\begin{result}[\Cref{thm:two-step-simulation}, One-prover reduction]
\label{res:sep-one}
For every integer \(k\ge 1\) and completeness--soundness parameters \(c,s\) satisfying \(c>(6+s)/7\),
\[
 \dQMAsep[k][c,s;\,p,r]
 \subseteq
 \dQMAsep\!\left[
 c,\frac{6+s}{7};\,3p+O(\log N),r+4
 \right].
\]
\end{result}

The reduction preserves completeness but weakens soundness. For protocols with perfect completeness, we achieve arbitrarily small soundness error with the same four-round overhead.

\begin{result}[\Cref{thm:perfect-amplification}, Amplification for \(\dQMAsep\)]
\label{res:sep-perfect}
For every integer \(\ell\ge1\) and constant \(s<1\),
 \[
 \dQMAsep{}[1,s;\,p,r]
 \subseteq
 \dQMAsep\!\left[
 1,2^{-\ell};\,
 O(\ell p+\log N),r+4
 \right].
\]
\end{result}
Beyond perfect completeness, we extend amplification to protocols on networks of bounded size, provided the initial completeness and soundness errors are sufficiently small.

Together, \Cref{res:sep-perfect} and the extension above show that one
prover suffices in both regimes. The local proof size is at most a
constant multiple of the original local proof size plus a logarithmic
term in the number of nodes, and the simulation uses four additional
rounds. In particular, for every integer \(k\ge 2\) and constant \(s<1\),
\[
  \dQMAsep{}[1,s]=\dQMAsep[k][1,s].
\]

\subsection{Overview of our techniques}\label{sec:techniques}

We summarize the main technical arguments behind our results.

\paragraph{Product testing without communication.}
We first reduce $\dQMAsep[k]$ to $\dQMAsep[2]$  by asking each of the two provers to send all $k$ original proofs. This may compromise the original protocol's soundness guarantee: while the two submitted blocks are unentangled, each prover may entangle the \(k\) registers sent to a node.

Harrow and Montanaro~\cite{HM13} address this obstacle in the centralized setting via the product test.
The two provers supply blocks $A=A_1\cdots A_k$ and $B=B_1\cdots B_k$, which are unentangled with each other.
Here, $A_j$ and $B_j$ correspond to the original proof from prover $j$. The product test applies a SWAP test to each pair $(A_j,B_j)$ and accepts if all $k$ tests accept. Because $A$ and $B$ are unentangled, these tests can detect entanglement among the registers within either block, for example, between \(A_i\) and \(A_j\). Their verifier chooses with equal probability between this product test and the original verification on one of the two blocks.

\begin{figure}
    \centering

\usetikzlibrary{arrows.meta}
\begingroup%
\definecolor{ptAmber}{HTML}{B87500}%
\definecolor{ptBlue}{HTML}{2367A0}%
\definecolor{ptInk}{HTML}{202A34}%
\definecolor{ptMuted}{HTML}{596570}%

\resizebox{\linewidth}{!}{%
\begin{tikzpicture}[
  x=1cm,y=1cm,
  reg/.style={
    draw=#1,fill=#1!22,line width=.7pt,
    rounded corners=1.3pt,
    minimum width=16mm,minimum height=6mm,
    inner sep=3pt,text=ptInk
  },
  test/.style={
    draw=ptInk!65,fill=white,line width=.7pt,
    rounded corners=1.3pt,
    minimum width=23mm,minimum height=15mm,
    inner sep=3pt,text=ptInk,font=\small
  },
  prover/.style={
    draw=#1,fill=white,line width=.7pt,
    rounded corners=1.3pt,
    minimum width=21mm,minimum height=7mm,
    inner sep=3pt,text=ptInk
  },
  wire/.style={draw=ptMuted,line width=.7pt},
  heading/.style={font=\small\bfseries,text=ptInk},
  note/.style={font=\footnotesize,text=ptMuted}
]

\fill[ptInk!10,rounded corners=3pt]
  (.15,2.70) rectangle (16.70,5.00);
\fill[ptInk!10,rounded corners=3pt]
  (.15,.15) rectangle (16.70,2.45);

\filldraw[
  draw=ptAmber!75,fill=ptAmber!7, fill opacity = 0.5,
  line width=.65pt,rounded corners=3pt
]
  (1.80,1.20) rectangle (3.90,4.80);
\filldraw[
  draw=ptBlue!75,fill=ptBlue!7, fill opacity = 0.5,
  line width=.65pt,rounded corners=3pt
]
  (4.15,.30) rectangle (6.25,3.90);

\node[prover=ptAmber] (proverA) at (2.85,5.65) {Prover $A$};
\node[prover=ptBlue] (proverB) at (5.20,5.65) {Prover $B$};
\draw[draw=ptAmber,line width=.8pt,
      -{Stealth[length=2.1mm,width=1.4mm]}]
  (proverA.south) -- (2.85,4.83);

\foreach \vertex/\ya/\yb/\yc in {
  u/4.25/3.35/3.80,
  v/1.75/.85/1.30
}{
  \node[heading] at (.95,\yc) {Node $\vertex$};

  \node[reg=ptAmber] (a\vertex) at (2.85,\ya)
    {$A_{j,\vertex}$};
  \node[reg=ptBlue] (b\vertex) at (5.20,\yb)
    {$B_{j, \vertex}$};

  \node[test] at (8.00,\yc)
    {\shortstack{SWAP test\\for each $j$}};

  \draw[wire,draw=ptAmber] (a\vertex.east) -- (6.85,\ya);
  \draw[wire,draw=ptBlue] (b\vertex.east) -- (6.85,\yb);

  \draw[wire,draw=ptAmber] (9.15,\ya) -- (12.50,\ya);
  \draw[wire,draw=ptBlue] (9.15,\yb) -- (9.65,\yb);
  \draw[wire] (9.65,\yb-.12) -- (9.65,\yb+.12);

  \node[note,anchor=west] at (9.85,\yb)
    {$B_{j, \vertex}$ unused};
}

\begin{scope}
  \clip (5.10,4.13) rectangle (5.30,4.37);
  \draw[ptInk!10,line width=3pt] (5.20,4.45) -- (5.20,4.05);
\end{scope}
\draw[draw=ptBlue,line width=.8pt,
      -{Stealth[length=2.1mm,width=1.4mm]}]
  (proverB.south) -- (5.20,3.93);

\filldraw[
  draw=ptInk!60,fill=white,
  rounded corners=3pt,line width=.7pt
]
  (12.50,.50) rectangle (16.50,4.60);

\node[heading,align=center] at (14.50,2.90)
  {Original distributed\\verification on $A$};
\node[note,align=center,text width=34mm] at (14.50,1.65)
  {Runs regardless of\\the test outcomes};

\end{tikzpicture}%
}%
\par
\endgroup%
    \caption{Reduction of a $\dQMAsep[k]$ protocol to $\dQMAsep[2]$ without additional communication.}
    \label{fig:distributed-product-test}
\end{figure}
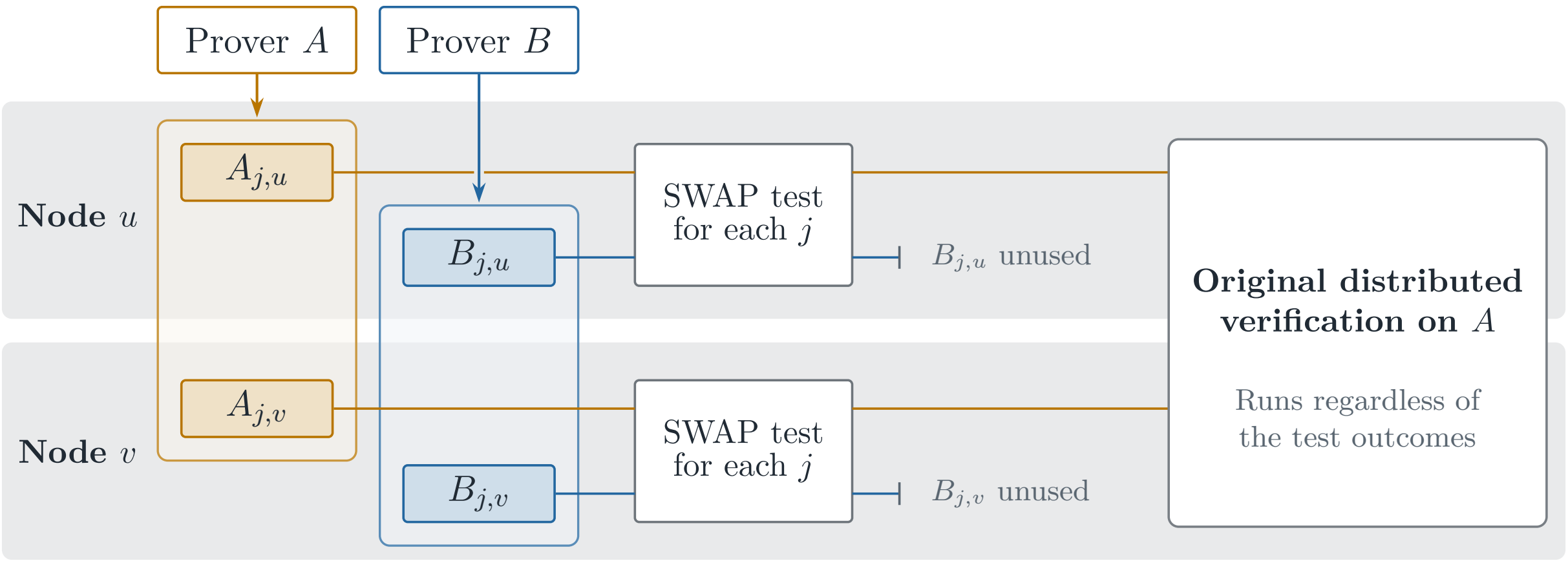

In our $\dQMAsep[2]$ setting, making the same random choice at every node would require coordination or shared randomness across the network. We avoid this by performing both stages sequentially, as illustrated in \Cref{fig:distributed-product-test}. The product test itself requires no communication. Each node $v$ receives one pair of registers for each original prover:
\[
  (A_{1,v},B_{1,v}),\quad (A_{2,v},B_{2,v}),\quad
  \ldots,\quad (A_{k,v},B_{k,v}).
\]
The separation between blocks $A$ and $B$ ensures that $A_{j,v}$ and $B_{j,v}$ are unentangled for every $j$.
Registers within the same block, such as $A_{i,v}$ and $A_{j,v}$, may nevertheless be entangled. To test for such entanglement, each node $v$ applies a SWAP test to every pair $(A_{j,v},B_{j,v})$. Together, these local tests implement the product test across the partition indexed by both the original provers and the nodes.

If any SWAP test rejects, node $v$ fixes its final output to reject but continues to participate in the protocol.
The nodes then run the original verification on block $A$.
Each node accepts only if all its local SWAP tests accept and it accepts in the original verification.

Since acceptance in $\dQMAsep[2]$ requires every node to accept, a failed SWAP test at any node causes global rejection. The nodes therefore enforce the product test's acceptance condition without exchanging their test outcomes. This yields a two-prover reduction without additional communication. \Cref{prot:product-test} gives the full construction and \Cref{lem:product-test} its proof of correctness.

\paragraph{Two unentangled blocks from one prover.}
Our earlier construction relies on two unentangled blocks $A$ and $B$. This guarantee is built into $\dQMAsep[2]$, but a single prover may entangle them. We recover it by exploiting the separability of $\dQMAsep$ proofs across nodes: each node receives its second block from a \emph{supplier node} that lies at distance at most two.

\begin{figure}[b]
  \centering
  \resizebox{\linewidth}{!}{%
    \begingroup%
\begin{tikzpicture}[
  x=1cm,y=1cm,
  reg/.style={
    draw=#1,fill=#1!22,line width=.7pt,
    rounded corners=1.3pt,
    minimum width=18mm,minimum height=6mm,
    inner sep=3pt,font=\normalsize,text=ptInk
  },
  heading/.style={font=\small\bfseries,text=ptInk},
  note/.style={font=\small,text=ptMuted},
  route/.style={
    -{Stealth[length=2.3mm,width=1.5mm]},
    line width=1.1pt
  }
]

\fill[ptInk!10,rounded corners=3pt]
  (.2,2.74) rectangle (18.4,4.94);
\fill[ptInk!10,rounded corners=3pt]
  (.2,.14) rectangle (18.4,2.34);

\draw[
  draw=black,fill=white,
  fill opacity=.55,draw opacity=1,
  line width=.7pt,rounded corners=3pt
]
  (2.55,-.02) rectangle (5.25,5.10);

\draw[
  draw=ptAmber!65,fill=ptAmber!7,
  line width=.6pt,rounded corners=3pt
]
  (2.75,2.97) rectangle (5.05,4.71);
\draw[
  draw=ptBlue!65,fill=ptBlue!7,
  line width=.6pt,rounded corners=3pt
]
  (2.75,.37) rectangle (5.05,2.11);

\node[note,text=black]
  at (3.9,2.54) {Unentangled};

\node[
  draw=black,fill=white,
  fill opacity=.55,draw opacity=1,text opacity=1,
  line width=.7pt,rounded corners=2pt,
  minimum height=6mm,inner sep=3pt,
  text=black
] (prover) at (3.9,5.82) {Single-prover};

\draw[route,draw=black]
  (prover.south) -- (3.9,5.15);

\draw[
  draw=ptAmber!75,fill=ptAmber!7, fill opacity = 0.5,
  line width=.65pt,rounded corners=3pt
]
  (10.80,1.18) rectangle (14.30,5.15);
\draw[
  draw=ptBlue!75,fill=ptBlue!7, fill opacity = 0.5,
  line width=.65pt,rounded corners=3pt
]
  (14.60,.26) rectangle (18.10,4.22);

\node[heading,text=ptAmber]
  at (12.55,4.90) {Global $A$};
\node[heading,text=ptBlue]
  at (16.35,3.97) {Global $B$};

\node[heading,anchor=west]
  at (.45,4.06) {Node $u$};
\node[note,anchor=west,text=ptAmber]
  at (.45,3.62) {\texttt{Amber}};
\node[heading,anchor=west]
  at (.45,1.46) {Node $v$};
\node[note,anchor=west,text=ptBlue]
  at (.45,1.02) {\texttt{Blue}};

\node[reg=ptAmber] (au0)
  at (3.9,4.30) {$A_u$};
\node[reg=ptAmber] (av0)
  at (3.9,3.38) {$A_v$};
\node[reg=ptBlue] (bu0)
  at (3.9,1.70) {$B_u$};
\node[reg=ptBlue] (bv0)
  at (3.9,.78) {$B_v$};

\node[reg=ptAmber,minimum width=24mm] (au1)
  at (12.55,4.30) {$A_u$};
\node[reg=ptBlue,minimum width=24mm] (bu1)
  at (16.35,3.38) {$B_u$};
\node[reg=ptAmber,minimum width=24mm] (av1)
  at (12.55,1.70) {$A_v$};
\node[reg=ptBlue,minimum width=24mm] (bv1)
  at (16.35,.78) {$B_v$};

\draw[route,draw=ptAmber]
  (au0.east) -- (au1.west);
\draw[route,draw=ptBlue]
  (bv0.east) -- (bv1.west);

\node[note]
  at (8.0,4.55) {kept at $u$};
\node[note]
  at (8.0,.53) {kept at $v$};

\draw[route,draw=ptAmber] (av0.east)
  .. controls (7.9,3.38) and (8.6,1.70)
  .. (av1.west);

\begin{scope}
  \clip (8.35,2.35) rectangle (8.75,2.55);
  \draw[draw=white,line width=3.6pt] (bu0.east)
    .. controls (8.4,1.70) and (9.0,3.38)
    .. (bu1.west);
\end{scope}

\draw[route,draw=ptBlue] (bu0.east)
  .. controls (8.4,1.70) and (9.0,3.38)
  .. (bu1.west);

\node[note,text=ptAmber]
  at (6.85,3.64) {$A_v:\ u\to v$};
\node[note,text=ptBlue]
  at (6.85,1.43) {$B_u:\ v\to u$};

\end{tikzpicture}%
\endgroup%
%
  }
  \caption{Nodes \(u\) and \(v\) prepare two unentangled blocks \(A, B\) from one prover in \(\dQMAsep\). Time flows from left (before the exchange) to the right (after the exchange).}
  \label{fig:routing}
\end{figure}
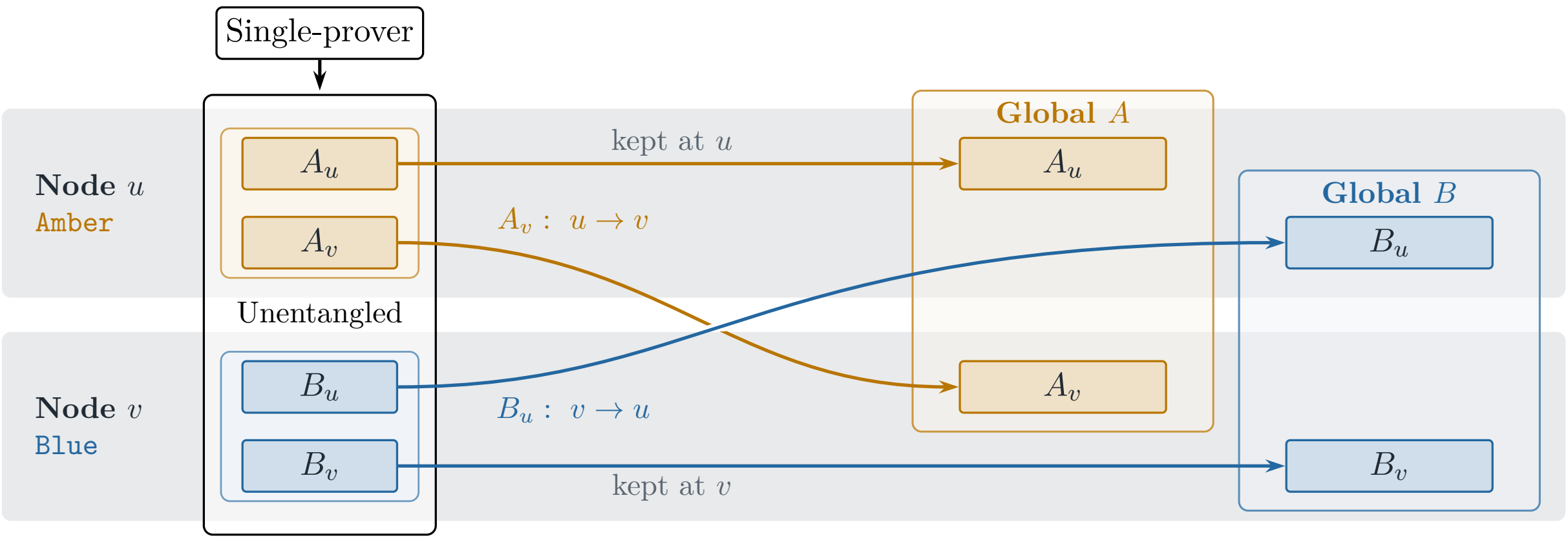

The idea is to partition the nodes into two colors: the prover sends block \(A\) to amber nodes and block \(B\) to blue nodes. Each node \(v\), say amber, receives its \(A\)-portion \(A_v\) directly from the prover, as well as the ID of its blue supplier \(u\). Supplier \(u\) receives from the prover both its own \(B\)-portion \(B_u\) and \(v\)'s portion \(B_v\), which it sends to \(v\) at the start of the protocol. By node-separability of $\dQMAsep$, this equips $v$ with a separable pair $(A_v,B_v)$ and, more strongly, ensures that the full blocks $A$ and $B$ are separable. Moreover, each node $v$ uses its color to distinguish $A_v$ from $B_v$, allowing the nodes to run the original $\dQMAsep[k]$ protocol on $A$. \Cref{fig:routing} illustrates an example where \(u\) and \(v\) reciprocally supply one another.

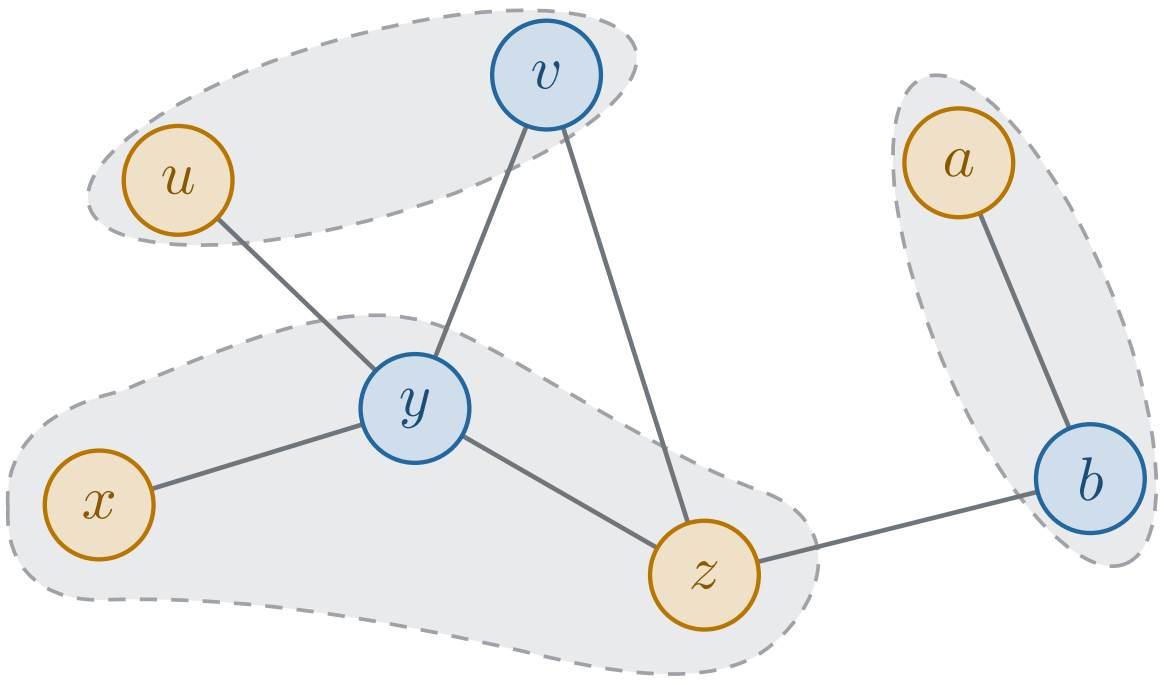
\begin{figure}[t]
  \centering
\begingroup
\definecolor{ptAmber}{HTML}{B87500}%
\definecolor{ptBlue}{HTML}{2367A0}%
\definecolor{ptInk}{HTML}{202A34}%
\providecommand{\RoutingPartitionWidth}{.485\linewidth}%
\resizebox{\RoutingPartitionWidth}{!}{%
\begin{tikzpicture}[
  x=1cm,y=1cm,
  vertex/.style={circle,minimum size=6.2mm,inner sep=0pt,
    line width=.7pt,font=\small},
  amber/.style={vertex,fill=ptAmber!22,draw=ptAmber,
    text=ptAmber!75!black},
  blue/.style={vertex,fill=ptBlue!22,draw=ptBlue,
    text=ptBlue!75!black},
  network/.style={draw=ptInk!65,line width=.75pt},
  group/.style={draw=ptInk!43,fill=ptInk!10,line width=.6pt,
    dash pattern=on 2.8pt off 2.3pt}
]
  \draw[group,rotate around={15.945:(1.25,2.65)}]
    (1.25,2.65) ellipse[x radius=1.62,y radius=.52];
  \draw[group]
    (-.77,.48)
    .. controls (-.80,.85) and (-.52,1.06) .. (-.12,1.15)
    .. controls (.80,1.55) and (1.30,1.70) .. (1.80,1.48)
    .. controls (2.35,1.22) and (2.95,.78) .. (3.52,.58)
    .. controls (3.98,.42) and (3.92,-.02) .. (3.58,-.30)
    .. controls (3.25,-.58) and (2.60,-.44) .. (2.15,-.33)
    .. controls (1.30,-.14) and (.60,-.02) .. (-.20,-.04)
    .. controls (-.56,-.06) and (-.79,.13) .. cycle;
  \draw[group,rotate around={-67.38:(5.025,1.55)}]
    (5.025,1.55) ellipse[x radius=1.50,y radius=.52];

  \coordinate (up) at (.20,2.35);
  \coordinate (vp) at (2.30,2.95);
  \coordinate (xp) at (-.25,.50);
  \coordinate (yp) at (1.55,1.05);
  \coordinate (zp) at (3.20,.10);
  \coordinate (ap) at (4.65,2.45);
  \coordinate (bp) at (5.40,.65);

  \draw[network] (up) -- (yp);
  \draw[network] (vp) -- (yp);
  \draw[network] (xp) -- (yp) -- (zp);
  \draw[network] (zp) -- (bp) -- (ap);
  \draw[network] (vp) -- (zp);

  \node[amber] (u) at (up) {$u$};
  \node[blue]  (v) at (vp) {$v$};
  \node[amber] (x) at (xp) {$x$};
  \node[blue]  (y) at (yp) {$y$};
  \node[amber] (z) at (zp) {$z$};
  \node[amber] (a) at (ap) {$a$};
  \node[blue]  (b) at (bp) {$b$};
\end{tikzpicture}%
}%
\endgroup%
  \caption{Supply chain on a general network. Recipients and suppliers have
distinct colors.}
  \label{fig:supply}
\end{figure}

The next step is to ensure that such a supply chain can be established on any network. We do so in \Cref{lem:two-step-routing-cut} by grouping the nodes into pairs and triples. The pairs \(\{u,v\}\) consist of nodes at distance at most two that supply one another, using an intermediate \emph{relay node} when necessary. In a triple \(x-y-z\), node \(y\) supplies both neighbors \(x,z\) and is supplied by one of them. \Cref{fig:supply} illustrates this construction. The prover can safely provide both the coloring and the supply chain: the nodes check that each supplier and its recipients have opposite colors and verify the local consistency of the supply chain. If any check fails at a node, the node rejects.

The full construction and its proof of correctness are given in \Cref{thm:two-step-simulation}.

\paragraph{Amplification.}
An obstacle to quantum error reduction is that verification may destroy the witness, compromising further repetitions of the protocol. Strong error reduction permits witness reuse in centralized single-prover protocols~\cite{MW05}. With multiple provers, however, amplification must also account for separability between their proofs. Here, weak error reduction uses $\ell$ witness copies from each prover, increasing proof size~\cite{HM13}. Our construction in \Cref{thm:perfect-amplification} implements weak error reduction for $\dQMAsep$, whose soundness guarantee still relies on separability across nodes.

In particular, suppose the single-prover \(A\) sends nodes \(u\) and \(v\) the following \(\ell\) registers each: 
\[A_u^{[1]}, \quad A_u^{[2]}, \quad \dots, \quad A_u^{[\ell]}; \qquad A_v^{[1]}, \quad A_v^{[2]}, \quad \dots, \quad A_v^{[\ell]}.\]
Nodes \(u,v\) repeat the verification on each copy \((A_u^{[i]}, A_v^{[i]})\), and accept if all repetitions accept.

By node-separability of \(\dQMAsep\), the first pair \((A_u^{[1]}, A_v^{[1]})\) is unentangled: \(u\) and \(v\) can safely run the original verification on this pair, with the original soundness guarantee. However, a malicious prover may entangle $A_u^{[1]}$ with $A_u^{[2]}$, and $A_v^{[1]}$ with $A_v^{[2]}$. Conditioned on acceptance, the first repetition may then transfer this entanglement to the second pair $(A_u^{[2]},A_v^{[2]})$, breaking the separability promise for the second repetition.

Harrow and Montanaro~\cite{HM13} prevent this entanglement transfer by using an accepting effect that is separable across the relevant partition---$A_u:A_v$ in the example above.\footnote{Here, separability refers to a measurement effect, rather than a witness state.}
Our sequential implementation of testing and verification need not have this property. Nevertheless, in \Cref{lem:separable-upper-bound} we show that its accepting effect admits a separable upper bound with the same soundness guarantee.
This suffices to reduce soundness exponentially, even when a dishonest prover entangles registers from different repetitions.

Amplification in $\dQMAsep$ comes with a second hurdle: taking a majority vote over the global outcomes of the repetitions can require substantial additional communication.
Our construction so far has each node accept if and only if all its repetitions accept. This preserves perfect completeness; for \(c < 1\), however, the completeness guarantee drops to $c^\ell$, which decreases exponentially with the number of repetitions. Our construction in \Cref{thm:constant-error-amplification} chooses a suitable threshold for the fraction of accepting repetitions at each node, allowing completeness and soundness errors to decrease when both initial errors are sufficiently small.

\subsection{Other related works}

Distributed quantum algorithms have been studied in terms of both round
and message complexity. In the quantum CONGEST model, algorithms are
known for computing or approximating diameter, radius, and
girth~\cite{LeGall+PODC18,vAdV22,WY22}, and for detecting triangles,
cliques, and even cycles~\cite{Izumi+STACS20,CensorHillel+22,FLMT24}.
In the quantum routing model~\cite{DMP25}, Dufoulon, Magniez, and
Pandurangan obtained nearly tight message complexity bounds for leader
election, broadcast, and minimum spanning tree
construction~\cite{Dufoulon+2026}, while Le Gall et al.\ showed an
exponential quantum advantage in message complexity for routing on
welded-tree networks~\cite{LeGall+Routing2025}.

In the quantum LOCAL model~\cite{GKM09}, Le Gall, Nishimura, and
Rosmanis exhibited tasks on $N$-node networks that are solvable in a
constant number of quantum rounds but require $\Omega(N)$ randomized
classical rounds~\cite{LNR19}. For locally checkable problems, Balliu
et al.\ first established a quantum advantage that grows with the
maximum degree~\cite{B+24} and later obtained an asymptotic advantage
on bounded-degree graphs~\cite{A+25}. In a recent breakthrough, Flin,
Li, and Suomela gave a constant-round quantum algorithm for
$3$-coloring cycles with high probability \cite{FLS26}.

There is also a substantial body of work on lower bounds on the round
complexity of distributed quantum
algorithms~\cite{EKNP14,AF14,GKZ19,LR22,CoiteuxRoy+24,Dhar+24,Akbari+25,Balliu+DISC2025,BrandtGottlicher2026,FMZ26,CoiteuxRoy+26,GL26}.

\subsection{Organization}

The paper is organized as follows. In \Cref{sec:model}, we fix notation and introduce the background needed for our results. In \Cref{sec:separable}, we present the constructions underlying \Cref{res:sep-two,res:sep-one} and prove their correctness.
In \Cref{sec:amplification-collapse}, we amplify these protocols to obtain \Cref{res:sep-perfect} and establish the resulting prover collapse for \(\dQMAsep[k]\). Finally, in \Cref{sec:no-comms}, we show how larger local proofs allow a reduction to a single prover without any communication overhead.

\subsection{Discussion}

Our results show that a single prover suffices for distributed quantum
Merlin--Arthur protocols with node-separable proofs and perfect completeness.
In the centralized setting, by contrast, recent quantum~\cite{BGHHHNN26}
and classical~\cite{GMT26} oracle separations between $\mathsf{QMA}$ and
$\mathsf{QMA}(2)$ provide evidence that two unentangled provers are more
powerful than one. We discuss two directions for further research.

\paragraph{Node-entangled proofs.} The \(\dQMA\) model relaxes the separability condition of \(\dQMAsep\), allowing each prover's witness to be entangled across nodes. Does a single prover still suffice, that is, does \(\dQMA[k]=\dQMA\) hold for \(k\ge2\)? Our distributed product test does not directly extend to this setting because completeness is no longer guaranteed.
While the local SWAP tests correctly test for entanglement among the \(k\) original proofs, they also penalize entanglement across nodes, which is allowed even in honest \(\dQMA\) proofs.

\paragraph{Strong error reduction.} Our amplification construction implements parallel repetition, causing the local proof size to scale linearly with the number of runs of the protocol. In $\mathsf{QMA}$, strong error reduction avoids this growth by alternating applications of the verifier circuit and its inverse~\cite{MW05}. In $\dQMAsep$, however, reversal after measurement need not restore separability across nodes, complicating the soundness analysis. Such an approach would also increase the round complexity linearly with the number of repetitions. Overcoming this obstacle could advance efforts to prove $\mathsf{QMA}(2)=\mathsf{NEXP}$~\cite{BFLMW26}.

\subsection{AI disclosure}

The authors derived \Cref{res:sep-two} without AI assistance. They also independently derived initial versions of \Cref{res:sep-one} and \Cref{res:sep-perfect} restricted to bipartite networks. During September and October 2026, GPT-6 Astra assisted in repairing the proofs and extending \Cref{res:sep-one} and \Cref{res:sep-perfect} to general networks.  These interactions yielded the technical arguments for amplification with imperfect completeness and those presented in \Cref{sec:no-comms}. The same model also helped the authors draft and revise the manuscript. The authors reviewed all technical arguments in full and take full responsibility for the mathematical claims and final text.

\subsection{Acknowledgments}

This work was supported in part by JSPS KAKENHI (Grant Nos.~JP24H00071,
JP24K22293, JP25K24465, JP25K24674, and JP26H02504), MEXT Q-LEAP
(Grant No.~JPMXS0120319794), JST ASPIRE (Grant No.~JPMJAP2302),
JST CREST (Grant No.~JPMJCR24I4), and JST SPRING (Grant No.~JPMJSP2125).
\section{Preliminaries}
\label{sec:model}

We assume familiarity with the basics of quantum computation~\cite{NC10}.
Throughout, we write $[k]=\{1,\ldots,k\}$, take logarithms to base two,
and work with finite-dimensional Hilbert spaces.

\subsection{Quantum notation}
\label{sec:quantum-notation}

An $n$-qubit register has Hilbert space $(\mathbb C^2)^{\otimes n}$.
A quantum state is a positive semidefinite operator of trace one.
A pure state is represented by a unit vector $\ket{\psi}$, identified with $\ket{\psi}\!\bra{\psi}$. We use $\|\cdot\|$ for the Euclidean norm of vectors. For a state $\rho_{AB}$, its reduced state on $A$ is $\rho_A=\operatorname{tr}_B\rho_{AB}$.
The relation $X\le Y$ means that $Y-X$ is positive semidefinite.
An \emph{effect} is an operator $0\le E\le I$; a measurement with accepting effect $E$ accepts $\rho$ with probability $\operatorname{tr}(E\rho)$.

A state on $\mathcal H_1\otimes\cdots\otimes\mathcal H_q$ is \emph{separable} across this partition if it can be written as
\[
 \rho=\sum_a\lambda_a\,
       \rho_{a,1}\otimes\cdots\otimes\rho_{a,q},
 \qquad \lambda_a\ge0,\qquad \sum_a\lambda_a=1,
\]
where each $\rho_{a,i}$ is a state on $\mathcal H_i$.
For two registers $A,B$, we denote this set by $\operatorname{Sep}(A:B)$. A separable pure state is a product state.
By convexity, the maximum acceptance probability over separable states is attained on a pure product state. Operations local to the factors, including conditioning on local measurement outcomes, preserve
separability.

\begin{definition}[Separable effects]
\label{def:separable-effect}
An effect $\Omega$ on $\mathcal H_A\otimes\mathcal H_B$ is
\emph{separable across $A:B$} if
\begin{equation}\label{eq:separable-effect}
 \Omega=\sum_a C_a\otimes D_a,
 \qquad C_a\ge0,\quad D_a\ge0,
\end{equation}
where $C_a$ and $D_a$ act on $A$ and $B$, respectively.
\end{definition}

We use two bounds relating states and acceptance probabilities.
For unit vectors $\ket{\phi},\ket{\psi}$ and any effect $E$,
\begin{equation}\label{eq:effect-overlap}
 \sqrt{\braket{\phi|E|\phi}}
 \le \sqrt{\braket{\psi|E|\psi}}
     +\sqrt{1-|\braket{\phi|\psi}|^2}.
\end{equation}
This follows by decomposing $\ket{\phi}$ into its components parallel and orthogonal to $\ket{\psi}$, and applying the triangle inequality to $\sqrt E\ket{\phi}$.
For mixed states, the trace distance $\frac12\|\rho-\sigma\|_1$, where $\|X\|_1=\operatorname{tr}\sqrt{X^\dagger X}$, satisfies
\[
 \bigl|\operatorname{tr}(E\rho)-\operatorname{tr}(E\sigma)\bigr|
 \le\tfrac12\|\rho-\sigma\|_1.
\]
Quantum channels are completely positive trace-preserving maps; they contract trace distance.

\subsection{Distributed quantum verification}
\label{sec:sep-model}
\label{sec:resources}

Let $G=(V,E)$ be a finite simple undirected network with $N=|V|$ and no isolated nodes. Each node \(v\) receives a local input \(x_v\), has a unique \(O(\log N)\)-bit identifier, and knows its neighbors' identifiers and the corresponding communication ports. We write $N(v)$ for the neighbors of $v$ and $\operatorname{dist}_G(u,v)$
for graph distance. The distributed input $x=(x_v)_{v\in V}$ has length $n=\sum_v |x_v|$.

The nodes collectively verify a property of the network and its inputs with the help of $k$ computationally unbounded provers, each knowing the network and all local inputs. Before verification, prover $j$ sends each node $v$ a quantum proof register with Hilbert space $\mathcal H_{j,v}$, whose size is prescribed by the protocol.
The full witness space is $\mathcal H=\bigotimes_{j=1}^k\bigotimes_{v\in V}\mathcal H_{j,v}$.

We extend the node-separable model of~\cite{HKN24} to $k$ unentangled provers: proofs from distinct provers are unentangled, and each prover's proof is separable across nodes. By convexity, it suffices to consider pure witnesses of the form
\begin{equation}\label{eq:sep-witnesses}
 \ket{\psi}
 =\bigotimes_{j=1}^k\bigotimes_{v\in V}\ket{\psi_{j,v}},
 \qquad \ket{\psi_{j,v}}\in\mathcal H_{j,v}.
\end{equation}
Thus, the registers supplied by different provers to the same node are also initially unentangled.

After receiving the proofs, the nodes execute synchronous communication rounds. In each round, they perform local quantum operations and exchange quantum messages with their neighbors. 
Local computation and workspace are unrestricted, and exact operations are allowed.
Auxiliary registers are initialized locally, with no prior shared randomness or entanglement. 
Each node outputs accept or reject;
the protocol accepts if and only if all nodes accept.

A distributed language $L$ is a set of instances $(G,x)$, where $x$ is
a distributed input on $G$.

\begin{definition}[Separable distributed quantum verification]
\label{def:sep-verification}
For an integer $k\ge1$ and $c,s\in[0,1]$, a language $L$ belongs to $\dQMAsep[k][c,s;\,p,r]$ if it admits a protocol of the above form using at most $p$ proof qubits at each node, summed over all provers, and at most $r$ synchronous communication rounds, such that:
\begin{itemize}
\item \textbf{Completeness.} For every $(G,x)\in L$, some admissible
witness makes all nodes accept with probability at least $c$.
\item \textbf{Soundness.} For every $(G,x)\notin L$, every admissible
witness makes all nodes accept with probability at most $s$.
\end{itemize}
\end{definition}

Thus, $p$ bounds the \emph{maximum local proof size}. It includes classical proof bits and excludes local workspace. 
The parameters may depend on $n$ and $N$. We write $\dQMAsep=\dQMAsep[1]$ for the single-prover model and call $c=1$ \emph{perfect completeness}.

\subsection{The product test}
\label{sec:quantum-primitives}

For equal-dimensional registers $A,B$, the operator $\mathrm{SWAP}$ exchanges their contents. The SWAP test is the projective measurement with accepting projector $(I+\mathrm{SWAP})/2$.

The product test of Harrow and Montanaro~\cite{HM13} acts on two blocks $A=A_1\cdots A_q$ and $B=B_1\cdots B_q$, each with space $\mathcal H=\bigotimes_{i=1}^q\mathcal H_i$. 
It applies a SWAP test to each pair $A_i,B_i$ and accepts if all tests accept. With $\mathrm{SWAP}_i$ exchanging $A_i$ and $B_i$, its accepting projector is
\begin{equation}\label{eq:product-test}
 \Pi=\bigotimes_{i=1}^q\frac{I+\mathrm{SWAP}_i}{2}.
\end{equation}
For unit vectors $\ket{\alpha},\ket{\beta}\in\mathcal H$, write $\operatorname{PT}(\alpha,\beta) =\braket{\alpha\otimes\beta|\Pi|\alpha\otimes\beta}$ for its acceptance probability. Identical product states pass unchanged.
$\Pi$ is separable across $A:B$~\cite[proof of Lemma~6]{HM13}. For $\ket{\psi}\in\mathcal H_1\otimes\cdots\otimes\mathcal H_q$,
we write $\omega_\psi$ for its maximum squared overlap with a product vector,
\[
  \omega_\psi =
  \max_{\substack{\ket{\phi_i}\in\mathcal H_i,\ \|\phi_i\|=1\\
                  i=1,\ldots,q}}
  \bigl|\langle\psi\mid\phi_1\otimes\cdots\otimes\phi_q\rangle\bigr|^2.
\]

\begin{lemma}[Product-test overlap bound]
\label{lem:pt-cross}
For unit vectors $\ket{\alpha},\ket{\beta}\in\mathcal H$, let
$\omega_\alpha$ and $\omega_\beta$ be their respective maximum squared
overlaps with unit product vectors in $\mathcal H_1\otimes\cdots\otimes\mathcal H_q$.
Then
\begin{equation}\label{eq:pt-overlap-deficit}
 (1-\omega_\alpha)+(1-\omega_\beta)
 \le4\bigl(1-\operatorname{PT}(\alpha,\beta)\bigr).
\end{equation}
\end{lemma}
\begin{proof}
The reduced-state identity of~\cite[Lemma~2]{HM13} and Cauchy--Schwarz give
\[
 \operatorname{PT}(\alpha,\beta)
 \le\sqrt{\operatorname{PT}(\alpha,\alpha)
          \operatorname{PT}(\beta,\beta)}
 \le\frac{\operatorname{PT}(\alpha,\alpha)
          +\operatorname{PT}(\beta,\beta)}{2}.
\]
By~\cite[Corollary~1.2]{BGJW26}, $\operatorname{PT}(\alpha,\alpha)\le(1+\omega_\alpha)/2$, and likewise for $\beta$; the case $q=1$ is immediate.
Substitution gives the claim.
\end{proof}

We also use the following multiplicativity property of separable effects for parallel repetition.
\begin{lemma}[{\cite[Lemma~10]{HM13}}]
\label{lem:separable-multiplicativity}
Let $\Omega$ be an effect separable across $A:B$. For every integer $\ell\ge1$, let $A^{[i]},B^{[i]}$ be copies of $A,B$ for $i\in[\ell]$.
Every state $\rho$ separable across $A^{[1]}\cdots A^{[\ell]}:B^{[1]}\cdots B^{[\ell]}$ satisfies
\begin{equation}\label{eq:separable-multiplicativity}
 \operatorname{tr}(\Omega^{\otimes\ell}\rho)
 \le\left(
   \max_{\tau\in\operatorname{Sep}(A:B)}
   \operatorname{tr}(\Omega\tau)
 \right)^\ell.
\end{equation}
\end{lemma}
\section{Prover reduction via distributed product test}\label{sec:separable}

We reduce the number of provers from $k$ to two using the distributed product test in \Cref{prot:product-test}.

Let $\mathcal P$ be a $\dQMAsep[k]$ protocol with witness
space $\mathcal H=\bigotimes_{j,v}\mathcal H_{j,v}$.
Its witnesses are separable across all factors
$\mathcal H_{j,v}$, indexed by prover $j=1,\ldots,k$ and node $v\in V$. In \Cref{prot:product-test}, node \(v\) receives registers \(A_v = A_{1,v}, \dots, A_{k,v}\) and \(B_v = B_{1,v}, \dots, B_{k,v}\). While separability across $A_v:B_v$ is guaranteed, the registers within either block may be entangled. The nodes test the separability promise of \(\mathcal{P}\) by the local SWAP tests in Step~1, whose joint accepting effect on global registers \(A=(A_v)_{v\in V}\) and \(B=(B_v)_{v\in V}\) is
\begin{equation}\label{eq:distr_PT}
    \Pi=\bigotimes_{j,v}\frac{I+\mathrm{SWAP}_{j,v}}2.
\end{equation}
By \eqref{eq:product-test}, Step~1 globally implements
the product test across the  factors $\mathcal H_{j,v}$. Then the nodes run the original verification on register \(A\).

The separability requirement across $A:B$ in
\Cref{prot:product-test} conforms to the two-prover
promise of $\dQMAsep[2]$, and thus yields the following reduction from \(k\) to two provers.

\begin{figure}[!t]
    \begin{protocol}{Sequential product-test simulation}\label{prot:product-test} 
    \textbf{Goal:} Simulate a \(\dQMAsep[k]\) protocol $\mathcal P$ with witness space \(\mathcal{H} = \bigotimes_{j,v} \mathcal{H}_{j,v}\).

    \textbf{Witness:} Separable registers \(A:B\), each over \(\mathcal{H}\). Node $v$ receives $A_{j,v},B_{j,v}$ for $1 \leq j \leq k$.
    \begin{enumerate}[leftmargin=*,label=\arabic*.]
        \item Each node $v$ SWAP tests $A_{j,v}$ against $B_{j,v}$ for every $j$.
        \item Execute $\mathcal P$ on $A$,
        regardless of the SWAP test outcomes.
        \item Each node accepts if all its SWAP tests and its verification accept.
    \end{enumerate}
    \end{protocol}
\end{figure}

\begin{theorem}\label{lem:product-test}
On any network of \(N\) nodes, for any $k\ge2$ and $0\leq (6 + s)/7<c\leq1$,
\[
 \dQMAsep[k][c,s;\,p,r]
 \subseteq
 \dQMAsep[2]\!\left[
 c,\frac{6+s}{7};\,2p,r
 \right].
\]
\end{theorem}

\begin{proof}
Fix a \(\dQMAsep[k]\) protocol \(\mathcal{P}\). We show correctness of \Cref{prot:product-test}. By convexity, we may restrict attention to pure witnesses.

Two honest provers supply identical copies of an honest
product witness, preparing identical pure states in
$A_{j,v}$ and $B_{j,v}$ for every $j,v$.
All SWAP tests accept with certainty and leave the copies
unchanged, so running $\mathcal P$ on \(A\)
accepts with probability at least $c$.

For soundness, let $\ket{\Psi}=\ket{\alpha}_A\otimes\ket{\beta}_B$ be an arbitrary witness.
Let $\ket{\phi_\alpha}$ be the closest  product state to
$\ket{\alpha}$, with
$\omega_\alpha=|\langle\phi_\alpha|\alpha\rangle|^2$
their squared overlap.
Define $\ket{\phi_\beta}$ and $\omega_\beta$ analogously.

By \eqref{eq:distr_PT}, \Cref{prot:product-test} requires
the product test to pass on \(A,B\), followed by acceptance of
$\mathcal P$ on $A$.
Our proof balances rejection in the two stages: high success
probability in the product test leaves both registers \(A\) and \(B\) close to a product state, allowing us to use the soundness of $\mathcal P$.

First observe that exchanging $A$ and $B$ leaves the
unnormalized post-test state unchanged:
\begin{equation}\label{eq:exchange}
 \Pi(\ket{\alpha}_A\otimes\ket{\beta}_B)
 =
 \Pi(\ket{\beta}_A\otimes\ket{\alpha}_B).
\end{equation}
The overall acceptance probability is therefore unchanged,
so we may assume $\omega_\alpha\ge\omega_\beta$.

Let $E$ be the accepting effect of $\mathcal P$, and $R$
its rejecting effect when run on $A$.
Denote by $r_{\mathrm{PT}}$ the probability that the
product test rejects, and by $r_{\mathrm{ver}}$ the
probability that it passes but the subsequent verification
rejects. Explicitly,
\begin{equation}\label{eq:defs_DPT}
 R=(I-E)_A\otimes I_B,\qquad
 r_{\mathrm{PT}}=\|(I-\Pi)\ket{\Psi}\|^2,\qquad
 r_{\mathrm{ver}}=\|\sqrt R\,\Pi\ket{\Psi}\|^2.
 \end{equation}
These two events partition overall rejection, whose
probability is therefore
$\delta=r_{\mathrm{PT}}+r_{\mathrm{ver}}$.

Using $\omega_\alpha\ge\omega_\beta$ and \eqref{eq:pt-overlap-deficit} from \Cref{lem:pt-cross}
we obtain
\begin{equation}\label{eq:prod_bound}
 1-\omega_\alpha
 \le \frac{(1-\omega_\alpha)+(1-\omega_\beta)}{2}
 \le 2r_{\mathrm{PT}}.
\end{equation}

Since $\ket{\phi_\alpha}$ is a product witness, soundness
of $\mathcal P$ implies that it is rejected with probability
at least $1-s$.
Applying \eqref{eq:effect-overlap} to $I-E$, we obtain
\begin{equation}\label{eq:arg_3.1}
 \sqrt{1-s}
 \le \sqrt{\langle\alpha|(I-E)|\alpha\rangle}
      +\sqrt{1-\omega_\alpha}
 = \|\sqrt R\,\ket{\Psi}\|+\sqrt{1-\omega_\alpha}.
\end{equation}
Decomposing
$\ket{\Psi}=\Pi\ket{\Psi}+(I-\Pi)\ket{\Psi}$ gives
\begin{align}
 \sqrt{1-s}
 &\le \|\sqrt R\,\ket{\Psi}\|+\sqrt{1-\omega_\alpha}
 && \text{(by \eqref{eq:arg_3.1})}\\
 &\le \|\sqrt R\,\Pi\ket{\Psi}\|
      +\|\sqrt R\,(I-\Pi)\ket{\Psi}\|
      +\sqrt{1-\omega_\alpha}
 && \text{(triangle inequality)}
 \label{eq:pt-triangle}\\
 &\le \sqrt{r_{\mathrm{ver}}}
      +\sqrt{r_{\mathrm{PT}}}
      +\sqrt{1-\omega_\alpha}
 && \text{(Def. of $r_{\mathrm{ver}}$ and
 $R \leq I$)}
 \label{eq:pt-contraction}\\
 &\le \sqrt{r_{\mathrm{ver}}}
      +(1+\sqrt2)\sqrt{r_{\mathrm{PT}}}
 && \text{(by \eqref{eq:prod_bound})}
 \label{eq:pt-rejection}\\
 &\le \sqrt{(4+2\sqrt2)
      (r_{\mathrm{PT}}+r_{\mathrm{ver}})}\le \sqrt{7\delta}
 && \text{(Cauchy--Schwarz).}
 \label{eq:pt-cauchy}
\end{align}
Thus $\delta\ge(1-s)/7$, so the overall acceptance
probability is $1-\delta\le(6+s)/7$.
\end{proof}

The correctness of \Cref{prot:product-test} rests on the separability promise across \(A : B\). This guarantee is lost upon moving to the single prover setting \(\dQMAsep\). To restore it, we exploit separability across nodes. For each node $v$, the prover supplies one block directly to $v$ and a second to its \emph{supplier} $\mathsf{s}(v)$,
which forwards it to $v$ at the start of the protocol.
We organize these transfers using the following routing
structure.

\begin{definition}\label{def:two-step-routing-cut}
A \emph{two-step routing cut} is a pair
$(\chi,\mathsf{s})$, where
$\chi:V\to\{\texttt{Amber},\texttt{Blue}\}$ is a coloring
and $\mathsf{s}:V\to V$ a \emph{supplier map} satisfying the following conditions for every $v\in V$:
\begin{enumerate}
    \item Its supplier is within distance two:
    $\operatorname{dist}_G(v,\mathsf{s}(v))\le2$;
    \item It has a different color from its supplier:
    $\chi(\mathsf{s}(v))\neq\chi(v)$;
    \item It supplies at most two nodes:
    $\left|\{u\in V:\mathsf{s}(u)=v\}\right|\le2$.
\end{enumerate}
\end{definition}

Condition (1) bounds the rounds needed to deliver each block.
Condition (2) and node-separability guarantee the desired separable cut
$A:B$: these blocks consist of registers
initially sent to amber and blue nodes, respectively.
Finally, condition (3) bounds each node's proof size.

The following lemma guarantees that our routing strategy can take place over any network.

\begin{lemma}\label{lem:two-step-routing-cut}
Every graph without isolated vertices admits a
two-step routing cut.
\end{lemma}
\begin{proof}
Choose a rooted spanning tree in each connected component.
Process the tree from leaves to root, with the invariant that
every processed subtree is partitioned into pairs except
possibly for its root. At a vertex $v$, pair the unpaired
child roots with one another. If one remains, pair it
with $v$; otherwise leave $v$ unpaired. Every pair is
at tree distance at most two.

After processing the tree, only its root $z$ can remain
unpaired. If this happens, choose a child $u$ of $z$.
By construction, $u$ is paired either with one of its
own children or with another child of $z$. Adding $z$
to that pair gives a triple forming a path of length
two. Thus, each component is partitioned into pairs
and at most one such triple.

For each pair $\{u,v\}$, give its vertices opposite
colors and set $\mathsf{s}(u)=v$ and $\mathsf{s}(v)=u$.
For each triple $x-y-z$, give $y$ the opposite color
from $x,z$ and set $\mathsf{s}(x)=\mathsf{s}(z)=y$
and $\mathsf{s}(y)=x$. Every node has a supplier of opposite color at distance
at most two, and every supplier serves at most two
recipients. Hence $(\chi,\mathsf{s})$ is a two-step
routing cut.
\end{proof}

\begin{figure}[!t]
\begin{protocol}{Two-step routing simulation}
\label{prot:two-step-routing}
\textbf{Goal:} Simulate a $\dQMAsep[k]$ protocol $\mathcal P$.

\textbf{Witness at node $v$.}
Routing labels $\chi(v),\mathsf{s}(v),\mathsf{r}(v)$ and registers $O_v$ and $R_u$ for all \( \mathsf{s}(u) = v\).

\begin{enumerate}[leftmargin=*,label=\arabic*.]
\item[] \textbf{Initialization.} Node \(v\) measures $\chi(v),\mathsf{s}(v)$ and $\mathsf{r}(v)$.
\item \textbf{Round 1.}
Each node $v$ checks that $\mathsf{r}(v)\in N(v)$
and $\mathsf{s}(v)\ne v$. If both checks pass,
it sends a request $\big(v,\chi(v),\mathsf{s}(v)\big)$ to relay $\mathsf{r}(v)$.

\item \textbf{Round 2.} If \(\mathsf{r}(v) \ne \mathsf{s}(v)\), \(\textsf{r}(v)\) checks that
\(\mathsf{s}(v)\) is a neighbor and forwards
the request. 
Supplier \(\mathsf{s}(v)\) checks if \(v\) has the opposite color and if it received at most two requests.  On failure, \(\mathsf{s}(v)\) fixes rejection.

\item \textbf{Round 3.}
Supplier \(\mathsf{s}(v)\) returns $(R_v,v)$ to the node from
which the request arrived, serving at most two requests.\footnote{We assume \(\mathsf{s}(v)\) can identify \(R_v\) among its proof qubits. The assumption can be removed by ordering its outgoing registers and including the size of \(R_v\) in the request, without affecting the resource bounds.}

\item \textbf{Round 4.}
For indirect routes, relays forward $R_v$ to $v$.
A missing $R_v$ causes rejection at $v$ and is replaced
by a locally prepared register of the prescribed size.

\item \textbf{Verification.}
Nodes assign registers according to
\eqref{eq:register_assignment} and run
\Cref{prot:product-test} to simulate \(\mathcal{P}\).
Each node accepts if all its preliminary checks
and this execution accept.
\end{enumerate}
\end{protocol}
\end{figure}

Our one-prover reduction uses \Cref{prot:two-step-routing}
and achieves the following bounds.

\begin{theorem}\label{thm:two-step-simulation}
On any network of $N$ nodes, for every integer $k\ge 1$ and every \(c>(6+s)/7\),
\begin{equation}\label{eq:two-step-parameters}
 \dQMAsep[k]\!\left[c,s;\,p,r\right]
 \subseteq
 \dQMAsep\!\left[
 c,\frac{6+s}{7};\,3p+O(\log N),r+4
 \right].
\end{equation}
\end{theorem}
\begin{proof} Fix a \(\dQMAsep[k]\) protocol \(\mathcal{P}\). We show correctness of the simulation in \Cref{prot:two-step-routing} and quantify its resource bounds.
The one-prover reduction in \Cref{prot:two-step-routing} uses a
two-step routing cut $(\chi,\mathsf{s})$ supplied by the
prover. Each node $v$ receives its color $\chi(v)$,
the IDs of its supplier $\mathsf{s}(v)$ and
relay $\mathsf{r}(v)$, its own witness register $O_v$,
and a register $R_u$ for each node $u$ it supplies.
For routes of length two, the relay is an intermediate node
between $v$ and its supplier; for direct routes,
$\mathsf{r}(v)=\mathsf{s}(v)$. All nodes measure their color and routing IDs in the computational basis at the start of the protocol.

During the first four rounds, the nodes check the routing
cut $(\chi,\mathsf{s})$ and deliver the proof registers
accordingly. If any check fails at node $v$, it fixes its
final output to reject but continues to participate in
the protocol. Conditioned on all checks passing, each node $v$ now holds two unentangled registers
$O_v$ and $R_v$, which it relabels as follows:
\begin{equation}\label{eq:register_assignment}
     (A_v,B_v)=
     \begin{cases}
      (O_v,R_v), & \text{if }\chi(v)=\texttt{Amber},\\
      (R_v,O_v), & \text{if } \chi(v)=\texttt{Blue}.
     \end{cases}
 \end{equation}
Write $A=(A_v)_{v\in V}$ and $B=(B_v)_{v\in V}$.
Every register $A_v$ was initially supplied to an amber
node, whereas every register $B_v$ was initially supplied
to a blue node. By the node-separability promise of \(\dQMAsep\), the resulting blocks $A$ and $B$ are separable across $A:B$.

Thus, Step~5 satisfies the separability requirement of
\Cref{prot:product-test}, and the soundness argument of \Cref{lem:product-test}
bounds its acceptance probability by $(6+s)/7$ on
no-instances, conditioned on all routing checks passing.
If any routing check fails at a node, that node, and hence
the protocol, rejects with certainty. The same soundness
bound therefore holds unconditionally.

For completeness, the prover sends a valid two-step
routing cut $(\chi,\mathsf{s})$, whose existence is
guaranteed by \Cref{lem:two-step-routing-cut},
with valid relay IDs. All routing checks succeed with certainty. 
The prover also sends two identical copies of an honest
witness in blocks $A$ and $B$ so that, after routing, each node $v$
holds the intended registers $A_v =A_{1,v},\ldots,A_{k,v}$ and $B_v =B_{1,v},\ldots,B_{k,v}$, with $A_{j,v}$ and
$B_{j,v}$ in identical pure states.
The completeness argument of \Cref{lem:product-test}
therefore applies, giving acceptance probability at least $c$.

It remains to bound the resources required by
\Cref{prot:two-step-routing}. The two-step routing strategy adds four rounds to the
\(r\) required by \Cref{prot:product-test}. Each register \(O_v\) or \(R_u\) takes \(p\) qubits, the proof size of \(\mathcal P\). Since \(v\) supplies at most two nodes, these registers use at most \(3p\) qubits in total. Since each node ID uses \(O(\log N)\) bits,
the routing labels require \(O(\log N)\) qubits in total.
\end{proof}

\section{Amplification and prover collapse}
\label{sec:amplification-collapse}

In this section, we amplify \Cref{prot:product-test}
to compensate for the weakened soundness guarantee
of our reductions. We first consider perfect completeness
and then almost-perfect completeness. Under the respective
error assumptions, we obtain the prover collapse to one.
\paragraph{Construction.} Our amplification construction in \Cref{prot:parallel-repetition} implements parallel repetition on top of the routing strategy of \Cref{prot:two-step-routing}.
To each node \(v\), the prover now sends $\ell$ copies of $v$'s own proof $O_v$ and of
each proof $R_u$ that $v$ supplies. We denote these blocks as
\begin{equation*}\label{eq:copy-registers}
 O_v=O_v^{[1]},\ldots,O_v^{[\ell]}, \qquad
 R_u=R_u^{[1]},\ldots,R_u^{[\ell]}, \quad \text{for } \mathsf{s}(u) = v.  
\end{equation*}

Conditioned on all routing checks passing at Step~1, the routing
in \Cref{prot:two-step-routing} and the relabeling in
\eqref{eq:register_assignment} yield global blocks
$A=(A_v)_{v\in V}$ and $B=(B_v)_{v\in V}$ that are
separable across $A:B$. Grouping their registers by copy,
we can write this partition as
\begin{equation}\label{eq:global_cut}
 A^{[1]}\cdots A^{[\ell]}:B^{[1]}\cdots B^{[\ell]},
\end{equation}
where $A^{[i]}=(A_v^{[i]})_{v\in V}$ and
$B^{[i]}=(B_v^{[i]})_{v\in V}$.

At Step~2, the nodes repeat \Cref{prot:product-test} on registers \(A^{[i]}, B^{[i]}\) for all \(i = 1, \dots, \ell\).

The main obstacle to proving soundness of our construction is \emph{entanglement swapping}.
After routing, all copies are separable across \(A^{[i]} : B^{[i]}\); this is crucial to preserve the soundness guarantee of \Cref{prot:product-test}. Nonetheless, a malicious prover may entangle registers from different copies, for instance $A^{[i]}$ with $A^{[j]}$ and $B^{[i]}$ with $B^{[j]}$. Conditioned on acceptance, the $i$-th repetition may
then entangle the $j$-th copy across $A^{[j]}:B^{[j]}$,
so the original soundness guarantee no longer applies
to the $j$-th repetition.

\begin{figure}[!t]
\begin{protocol}{Parallel repetition}
\label{prot:parallel-repetition}
\textbf{Goal.} Amplify a $\dQMAsep[k]$ protocol
$\mathcal P$ with one prover through $\ell\ge1$ parallel repetitions.

\textbf{Witness at node $v$.} Routing labels $\chi(v),\mathsf{s}(v),\mathsf{r}(v)$
and proof registers $O_v = O_v^{[1]},\ldots,O_v^{[\ell]}$ and $R_u = R_u^{[1]},\ldots,R_u^{[\ell]}$ for all \(\mathsf{s}(u) = v\).

\begin{enumerate}[leftmargin=*,label=\arabic*.]
\item[] \textbf{Initialization.}
Node $v$ measures $(\chi(v),\mathsf{s}(v),\mathsf{r}(v))$
in the computational basis.

\item \textbf{Routing.}
Execute Rounds~1--4 of \Cref{prot:two-step-routing} to route \(R_v\).

\item \textbf{Verification.}
For each $i=1,\ldots,\ell$, node $v$ assigns
$O_v^{[i]},R_v^{[i]}$ to $A_v^{[i]},B_v^{[i]}$
according to \eqref{eq:register_assignment}.
The nodes run \Cref{prot:product-test} in parallel
on all $\ell$ repetitions.
Each node accepts if and only if all its routing
checks pass and it accepts in every repetition.
\end{enumerate}
\end{protocol}
\end{figure}

Harrow and Montanaro~\cite{HM13} show that this
entanglement transfer cannot occur when the accepting
effect of each repetition is separable across
$A^{[i]}:B^{[i]}$.\footnote{Here, separability refers
to a measurement effect rather than a state;
see \Cref{def:separable-effect}.}
The accepting effect of \Cref{prot:product-test}
need not have this property. For our analysis, however,
it suffices to upper-bound it by a separable effect
$\Omega$ with the same soundness bound.
\Cref{lem:separable-upper-bound} provides such an effect.

\begin{lemma}\label{lem:separable-upper-bound}
Let $\Lambda$ be the accepting effect of
\Cref{prot:product-test} to simulate a protocol \(\mathcal{P}\) with soundness $s<1$.
There exists an effect $\Omega$, separable across $A:B$,
with $\Lambda\le\Omega$ and, on every no-instance,
\begin{equation}\label{eq:sep_bound}
 \max_{\tau\in\operatorname{Sep}(A:B)}
 \operatorname{tr}(\Omega\tau)\le\frac{6+s}{7},
\end{equation}
where $\operatorname{Sep}(A:B)$ denotes the set of
states separable across $A:B$.
\end{lemma}

\begin{proof}
Let $E$ be the accepting effect of the simulated protocol
$\mathcal P$. We construct our candidate $\Omega$ by
modifying the separable verifier of Harrow and
Montanaro~\cite{HM13}:
(a) with probability $4/7$, perform the product test on $A,B$;
(b) with probability $2/7$, verify a uniformly chosen block;
and (c) with probability $1/7$, accept unconditionally.
Write $\Pi$ and $Q$ for the accepting effects of
branches~(a) and~(b), respectively. Then
\[
 Q=\frac{E_A\otimes I_B+I_A\otimes E_B}{2},
 \qquad
 \Omega=\frac{I+4\Pi+2Q}{7}.
\]
Separability of $Q$ across $A:B$ is immediate from
\Cref{def:separable-effect}, while separability of $\Pi$ follows
from~\cite[proof of Lemma~6]{HM13}.
As a convex combination of separable effects,
$\Omega$ is separable.

Fix a no-instance and an arbitrary product witness
$\ket{\Psi}=\ket{\alpha}_A\otimes\ket{\beta}_B$.
Let $r_\mathrm{PT}$ and $r_Q$ denote the rejection
probabilities conditional on choosing branches~(a)
and~(b), respectively:
\[
 r_\mathrm{PT}=\langle\Psi|(I-\Pi)|\Psi\rangle,\qquad
 r_Q=\langle\Psi|(I-Q)|\Psi\rangle.
\]
Applying \eqref{eq:arg_3.1}
from the proof of \Cref{lem:product-test} to both registers \(A\) and \(B\) gives
\begin{equation*}
 \sqrt{1-s}\le \sqrt{\langle\alpha|(I-E)|\alpha\rangle}
      +\sqrt{1-\omega_\alpha},\qquad
 \sqrt{1-s}
 \le \sqrt{\langle\beta|(I-E)|\beta\rangle}
      +\sqrt{1-\omega_\beta},
\end{equation*}
where \(\omega_\chi\) is the maximum squared overlap of \(\ket{\chi}\) with a product state.
Next, we average these two inequalities and apply Cauchy--Schwarz to each of the two averages, giving
\begin{equation}
     \sqrt{1-s}
 \le \sqrt{r_Q}
    +\sqrt{\frac{(1-\omega_\alpha)+(1-\omega_\beta)}2}
 \le \sqrt{r_Q}+\sqrt{2r_\mathrm{PT}}.
\end{equation}
Here, the first inequality also uses the definition of
$r_Q$, while the second follows from
\eqref{eq:pt-overlap-deficit}. 
Squaring the last bound and using
$(x+y)^2\le2x^2+2y^2$ gives
\(
 1-s
 \le 4r_\mathrm{PT}+2r_Q.
\)

Since rejection occurs only through branches (a) and (b), the acceptance probability is
\[
 \braket{\Psi|\Omega|\Psi}
 =1-\frac{4r_\mathrm{PT}+2r_Q}7
 \le 1-\frac{1-s}{7}
 =\frac{6+s}{7}.
\]
By convexity, the bound holds for any state \(\tau\) separable across \(A:B\).

Finally, we prove $\Lambda\le\Omega$. For an arbitrary state \(\ket{\Psi}\), we compare the rejection probabilities:
\begin{equation*}
 \text{Pr\,[\Cref{prot:product-test} fails]} = 1-\braket{\Psi|\Lambda|\Psi}
 =r_\mathrm{PT}+r_{\mathrm{ver}}, \qquad
 1-\braket{\Psi|\Omega|\Psi}
 =\frac{4r_\mathrm{PT}+2r_Q}7.
\end{equation*}
As in \eqref{eq:defs_DPT}, $r_{\mathrm{ver}} =\langle\Psi|\Pi\bigl((I-E)_A\otimes I_B\bigr)\Pi|\Psi\rangle
$ is the probability that the product test passes but verification rejects. Since \(\ket{\Psi}=\Pi\ket{\Psi}+(I-\Pi)\ket{\Psi}\), the triangle inequality gives
\[
\begin{aligned}
 \sqrt{r_Q}
 =\|\sqrt{I-Q}\ket{\Psi}\|\le\|\sqrt{I-Q}\Pi\ket{\Psi}\|
     +\|\sqrt{I-Q}(I-\Pi)\ket{\Psi}\| \leq \sqrt{r_{\mathrm{ver}}}+\sqrt{r_{\mathrm{PT}}}.
\end{aligned}
\]
Here, the last inequality uses $I-Q\le I$ to get \(\sqrt{r_\mathrm{PT}}\)
and symmetry of the post-product test state, which gives
$\Pi(E\otimes I)\Pi=\Pi Q\Pi$ and hence
$\|\sqrt{I-Q}\Pi\ket{\Psi}\|^2=r_{\mathrm{ver}}$.

Squaring the bound, we obtain
\[
\begin{aligned}
 1-\braket{\Psi|\Omega|\Psi}
 &=\frac{4r_{\mathrm{PT}}+2r_Q}{7}\\
 &\le\frac{6r_{\mathrm{PT}}+2r_{\mathrm{ver}}
       +4\sqrt{r_{\mathrm{PT}}r_{\mathrm{ver}}}}7\\
 &\le r_{\mathrm{PT}}+r_{\mathrm{ver}}
 =1-\braket{\Psi|\Lambda|\Psi}
 && \text{(by $4\sqrt{ab}\le a+4b$)}.
\end{aligned}
\]
This comparison holds for every
state $\ket{\Psi}$, so $\Lambda\le\Omega$.
\end{proof}

Amplification for perfect completeness has each node
accept if and only if all its $\ell$ repetitions accept.
The separable upper bound of \Cref{lem:separable-upper-bound}
ensures that soundness decreases exponentially in $\ell$,
despite possible entanglement between repetitions. \Cref{thm:perfect-amplification} combines amplification
with the reduction from $k$ provers to one; setting
$k=1$ gives amplification for $\dQMAsep$.

\begin{theorem}[Amplification for perfect completeness]
\label{thm:perfect-amplification}
On any network of \(N\) nodes, 
for all integers $k,\ell\ge1$ and $s<1$,
\begin{equation}\label{eq:parallel-parameters}
 \dQMAsep[k][1,s;\,p,r]
 \subseteq
 \dQMAsep[1]\!\left[
 1,\left(\frac{6+s}{7}\right)^\ell;\,
 3\ell p+ O(\log N),r+4
 \right].
\end{equation}
\end{theorem}

\begin{proof}
For an arbitrary \(\dQMAsep[k]\) protocol \(\mathcal{P}\) we show correctness of \Cref{prot:parallel-repetition}.

For completeness, the prover supplies valid routing labels
and $\ell$ copies of the honest witness pair
from \Cref{prot:two-step-routing}. The copies are independent, so the same completeness argument applies to every repetition: since $c=1$, all
routing checks and repetitions accept with certainty.

For soundness, assume all routing checks at Step~1 accept. By \eqref{eq:global_cut}, the routed state $\rho$ is separable across the two blocks $A$ and $B$, each
spanning all $\ell$ repetitions. Let $\Lambda$ be the accepting effect of
\Cref{prot:product-test} and $\Omega$ its separable
upper bound from \Cref{lem:separable-upper-bound}.
The probability that all \(\ell\) repetitions of \Cref{prot:product-test} accept at Step~2 is then
\begin{equation}\label{eq:bound_multiple_reps}
\begin{aligned}
 \operatorname{tr}\!\left(\Lambda^{\otimes\ell}\rho\right)
 &\le \operatorname{tr}\!\left(\Omega^{\otimes\ell}\rho\right)
 && \text{($0\le\Lambda\le\Omega$)}\\
 &\le \left(
 \max_{\tau\in\operatorname{Sep}(A^{[i]}:B^{[i]})}
 \operatorname{tr}(\Omega\tau)
 \right)^\ell
 && \text{(by \Cref{lem:separable-multiplicativity})}\\
 &\le \left(\frac{6+s}{7}\right)^\ell
 && \text{(by \Cref{lem:separable-upper-bound})},
\end{aligned}
\end{equation}
which establishes the claimed bound.

For the resource bounds, the \(\ell\) repetitions of
\Cref{prot:product-test} run in parallel and thus still take \(r\) synchronous rounds.
Two-step routing adds four rounds.  Each register \(O_v^{[i]}\) or \(R_u^{[i]}\) takes \(p\) qubits. Including the routing labels, the local proof size is
at most \(3\ell p+O(\log N)\) qubits.
\end{proof}

When \(c<1\), requiring all \(\ell\) repetitions to accept in
\Cref{prot:parallel-repetition} gives a completeness guarantee of only
\(c^\ell\). A global threshold vote among the nodes would fix the problem at the price of a
higher round complexity. Instead, we use a local acceptance rule to give amplification whenever completeness and soundness errors are both sufficiently small.

\begin{theorem}[Amplification for small networks]
\label{thm:constant-error-amplification}
On any network of \(N\) nodes
where each node knows \(N\), for every \(0<\varepsilon<1/2\)
there exists an integer
\(\ell=O(N\log(1/\varepsilon))\)
such that, for every integer \(k\ge1\) and \(c,s\) satisfying
\(c-s\ge1-1/(14N)\),
\[
 \dQMAsep[k][c,s;\,p,r]
 \subseteq
 \dQMAsep[1]\!\left[
 1-\varepsilon,\varepsilon;\,
 3\ell p+O(\log N),r+4
 \right].
\]
\end{theorem}
In particular, when the network size \(N\) is polynomially bounded in $n$,
polynomially many repetitions \(\ell\) suffice to achieve
$\varepsilon=2^{-\operatorname{poly}(n)}$.
We use the Bernoulli case of Hoeffding's lemma in the following form.
For every $q\in[0,1]$ and $t>0$,
\begin{equation}\label{eq:hoeffding-bernoulli}
  1-q+qe^t \le \exp\!\left(qt+\frac{t^2}{8}\right).
\end{equation}

\begin{proof}
We modify the construction from \Cref{prot:parallel-repetition}
at Step~2 so that each node accepts if and only if
all its routing checks pass and at most
\(d:=\lfloor\ell/(8N)\rfloor\) of its \(\ell\) runs of
\Cref{prot:product-test} reject. Let \(X_i\) indicate global acceptance of
the \(i\)-th repetition with \(S=\sum_{i=1}^{\ell}X_i\).

For completeness, the prover supplies valid routing labels
and independent honest copies for the repetitions.
All routing checks pass with certainty, and each repetition
accepts globally with probability at least \(c\).
Since \(1-c\le1/(14N)\), the expected number of globally
rejecting repetitions can be bounded as \(\mathbb{E}[\ell -S] \leq \ell(1-c) \leq \ell/(14N)\).

If a node rejects the amplified protocol, the modified acceptance
rule implies that it rejects in at least \(d+1>\ell/(8N)\)
repetitions. Each of these repetitions also rejects globally,
since global acceptance requires every node to accept. A Chernoff bound therefore gives
\begin{equation}\label{eq:bound_comp}
 \Pr[\text{global rejection}]
 \le
 \Pr\!\left[\ell-S>\frac{\ell}{8N}\right]
 =
 \Pr\!\left[\ell-S> \left(1 +\frac{3}{4}\right)\frac{\ell}{14N}\right] \le
 \exp\!\left(-\frac{9\ell}{616N}\right),
\end{equation}
where the last step uses \(\mathbb{E}[\ell - S] \leq \ell/(14N)\).

For soundness, we first determine how many accepting repetitions are required for the amplified protocol to accept. If all nodes accept the amplified protocol, each rejects
at most \(d\) repetitions. Every globally rejecting repetition has
a rejecting node, so at most \(Nd\) repetitions reject globally. Hence, acceptance of the amplified protocol requires
\begin{equation}\label{eq:bound_count}
 S \geq \ell-Nd \geq \frac{7\ell}{8}.
\end{equation}
It remains to bound \(\Pr[S\ge 7\ell/8]\). However,
since the prover may entangle the copies, the repetitions need not
be independent, so a Chernoff bound does
not directly apply.

As in the correctness proof of \Cref{prot:two-step-routing},
assume all routing checks pass, in which case the
routed state is separable across \(A:B\).
For any subset \(J\subseteq[\ell]\), the argument of
\eqref{eq:bound_multiple_reps} bounds the probability that
all repetitions in \(J\) accept globally as follows:
\begin{equation}\label{eq:bound_ac}
 \mathbb{E}\!\left[\prod_{i\in J}X_i\right]
 =
 \Pr[X_i=1\text{ for all }i\in J]
 \le \left(\frac{6+s}{7}\right)^{|J|}.
\end{equation}
Consequently, for every \(t>0\) and using \(q = (6+s)/7\) for convenience,
\begin{equation}\label{eq:moment_bound}
    \begin{aligned}
     \mathbb{E}[e^{tS}]
     &=
     \mathbb{E}\!\left[
       \prod_{i=1}^{\ell}\bigl(1+(e^t-1)X_i\bigr)
     \right]\\
     &=
     \sum_{J\subseteq[\ell]}(e^t-1)^{|J|}
     \mathbb{E}\!\left[\prod_{i\in J}X_i\right]\le (1-q+qe^t)^\ell,
    \end{aligned}
\end{equation}
where the last inequality uses \eqref{eq:bound_ac} and the
binomial theorem. 
Even though the repetitions need not be independent,
\eqref{eq:moment_bound} allows us to use a Chernoff argument
to bound the global acceptance probability in \eqref{eq:bound_count}:
\begin{equation}\label{eq:bound_sound}
\begin{aligned}
 \Pr[\text{global acceptance}]
 &\le \Pr\!\left[S\ge\frac{7\ell}{8}\right]
 && \text{(by \eqref{eq:bound_count})}\\
 &\le e^{-7t\ell/8}\,\mathbb{E}[e^{tS}]
 && \text{(Markov's inequality)}\\
 &\le e^{-7t\ell/8}(1-q+qe^t)^\ell
 && \text{(by \eqref{eq:moment_bound})}\\
 &\le \exp\!\left[\ell\left(\frac{t^2}{8}
                  -t\left(\frac78-q\right)\right)\right]
 && \text{(by \eqref{eq:hoeffding-bernoulli})}\\
 &= \exp\!\left[-2\ell\left(\frac78-q\right)^2\right]
 && \text{(taking $t=4(7/8-q)>0$)}\\
 &\le \exp\!\left[-2\ell\left(\frac{5}{392}\right)^2\right]
 && \text{(since $q\le 169/196$).}
\end{aligned}
\end{equation}
Since $N\ge 2$, we have $s\le 1/(14N)\le 1/28$, and hence
$q=(6+s)/7\le 169/196$. Thus $7/8-q\ge 5/392>0$, which justifies
the choice of $t$ and the final inequality.
Since rejecting routing checks cause global rejection, the bound holds unconditionally.

Choosing \(\ell=\lceil C N\ln(1/\varepsilon)\rceil\)
for a sufficiently large constant \(C\) makes both
error bounds in \eqref{eq:bound_comp} and \eqref{eq:bound_sound}
at most \(\varepsilon\). Since only the local acceptance rule
changes, the resource bounds remain those of
\Cref{prot:parallel-repetition}.
\end{proof}

Combining our prover reduction with the preceding amplification
results yields the following collapse to a single prover.

\begin{corollary}[Prover collapse]
\label{cor:prover-collapse}
For every integer \(k\ge2\) and constant \(0 < \varepsilon<1\),
the following equalities hold, allowing a constant-factor
increase and an additive \(O(\log N)\) term in local proof size,
and four additional rounds:
\begin{enumerate}
 \item For protocols with perfect completeness,
 \[
  \dQMAsep[k][1,\varepsilon]
  =\dQMAsep[1][1,\varepsilon].
 \]
 \item On networks of \(N\le 1/(28\varepsilon)\) nodes,
 where each node knows \(N\),
 \[
  \dQMAsep[k][1-\varepsilon,\varepsilon]
  =\dQMAsep[1][1-\varepsilon,\varepsilon].
 \]
\end{enumerate}
\end{corollary}

\begin{proof}
For perfect completeness, apply
\Cref{thm:perfect-amplification} with a constant \(\ell\) such that
$
 \left((6+\varepsilon)/7\right)^\ell\le\varepsilon
$.
For the second claim, \(N\le1/(28\varepsilon)\) gives completeness--soundness gap at least $1-1/(14N)$, so
\Cref{thm:constant-error-amplification} applies with
$\ell=O(N\log(1/\varepsilon))=O(1)$.
Both simulations use local proof size \(O(p+\log N)\)
and \(r+4\) rounds. Ignoring all but one prover gives the reverse inclusions.
\end{proof}
\section{Prover reduction without additional communication}
\label{sec:no-comms}

Our preceding reductions to one prover use witness routing. When the local prover registers are
short, we can instead perform the reduction locally,
avoiding any communication overhead.

Our construction relies on properties of the symmetric subspace
$\operatorname{Sym}^L(\mathbb C^d)$, which consists of all vectors in
$(\mathbb C^d)^{\otimes L}$ that are invariant under every permutation
of the $L$ tensor factors. In particular, we use the following form of
Chiribella's quantum de Finetti theorem.

\begin{lemma}[{\cite[Theorem~2]{C11}}]
\label{lem:chiribella-definetti}
Let $B_1,\ldots,B_L$ be $d$-dimensional registers. There exists a
measure-and-prepare channel $\mathcal M$ from
$\operatorname{Sym}^L(\mathbb C^d)$ to $B_1$ such that, for every
reference register $R$ and every state $\rho$ supported on
$\mathcal H_R\otimes\operatorname{Sym}^L(\mathbb C^d)$, the state
$\sigma_{RB_1}=(\operatorname{id}_R\otimes\mathcal M)(\rho)$ is
separable across $R:B_1$ and satisfies
\begin{equation}\label{eq:chiribella-definetti}
 \tfrac12\|\rho_{RB_1}-\sigma_{RB_1}\|_1\le\frac{d}{L+d},
\end{equation}
where $\rho_{RB_1}$ is obtained by discarding $B_2,\ldots,B_L$.
\end{lemma}

In our one-prover reduction, for each original prover \(j\), node \(v\) expects \(L\) copies \(\ket{\psi_{j,v}}^{\otimes L}\) of its proof on registers \(B_1,\dots, B_L\); \(v\) retains \(B_1\) and discards the other copies. In the yes case, \(v\) recovers \(\ket{\psi_{j,v}}\) exactly. In contrast, the single dishonest prover may entangle \(B_1\) with the other registers by some state \(\rho\). \Cref{lem:chiribella-definetti} states that, if \(\rho\) lies in the symmetric subspace and the number of copies \(L\) is large enough, once \(B_2, \dots, B_L\) are discarded, \(B_1\) and the rest of the witness are close to a separable state. Repeating the argument for each encoded block $j$ at every node $v$ shows that the joint state of the $Nk$ retained proof registers is close to a witness that is fully separable across nodes and provers. The original $\dQMAsep[k]$ soundness bound therefore holds up to a small additive error. The actual construction repeats the argument for \(k-1\) provers: if \(k-1\) registers are separable, the remaining \(k\)-th register must also be.

One thing to address is that \(\rho\) need not satisfy the
symmetric-subspace assumption of \Cref{lem:chiribella-definetti}.
To enforce this condition, fix an isometry from
\(\operatorname{Sym}^L(\mathbb C^d)\) into a qubit register,
and call its image the valid encoding subspace.
To decode, a node measures whether the received register lies in
this subspace. On a valid outcome, it applies the inverse isometry;
otherwise, it discards the register and prepares a fixed symmetric
state.

\begin{theorem}\label{thm:short-witnesses}
On any network of \(N\) nodes, consider a
\(\dQMAsep[k]\) protocol with parameters \((c,s;\,p,r)\),
where \(k\ge2\).
Suppose that, at each node, all but one of the \(k\) proof
registers contain at most \(b\) qubits, and let \(p_A\)
bound the proof size of the remaining register at any node.
For every \(0<\eta<c-s\),
\begin{equation}\label{eq:short-witnesses}
 \dQMAsep[k]\!\left[c,s;\,p,r\right]
 \subseteq
 \dQMAsep\!\left[c,s+\eta;\,p',r\right],
\end{equation}
where
\[
 p'\le p_A+
 O\!\left(k\,2^b\log\frac{Nk}{\eta}\right).
\]
\end{theorem}

In particular, if $b=O(\log n)$ and $p_A$, $N$, $k$,
and $\log(1/\eta)$ are polynomially bounded in $n$,
then \(p'\) also is. At node $v$, let \(A_v\) denote the unencoded register, of size at most \(p_A\).

\begin{proof}
At each node \(v\), the construction replaces each of the
\(k-1\) registers other than \(A_v\) by a block of \(L\)
registers, each of dimension \(d=2^b\). By the above argument, the state of each
block is supported on \(\operatorname{Sym}^{L}(\mathbb C^d)\).
Let \(D\) denote the dimension of this subspace and \(q\)
the number of qubits used to encode each block. We set
\[
 L=\left\lceil\frac{dNk}{\eta}\right\rceil,\qquad
 D=\binom{L+d-1}{d-1},\qquad
 q=\lceil\log D\rceil.
\]

To each node \(v\), the prover sends register \(A_v\) and $k-1$  encodings.
Each node decodes them into the symmetric subspace,
replacing invalid encodings by a fixed symmetric state.
From each of the \(k-1\) decoded blocks, the node keeps one register and discards the other \(L-1\). Finally, it runs its part of \(\mathcal P \) on the \(k-1\) retained registers together with \(A_v\).

For completeness, choose an honest witness $|\psi\rangle$
as in \eqref{eq:sep-witnesses}.
At each node $v$, the prover sends $A_v$ in its honest
state and encodings of $|\psi_{j,v}\rangle^{\otimes L}$
for the other $k-1$ original registers.
Decoding and keeping one register from each block
recover $|\psi\rangle$ exactly, preserving completeness $c$.

For soundness, each decoded block is supported on
$\operatorname{Sym}^{L}(\mathbb C^d)$, even for invalid
encodings, which the decoder replaces by a fixed
symmetric state. Consider one decoded block $B_1,\ldots,B_L$, and let
$R$ contain all registers outside it. The actual protocol
discards $B_2,\ldots,B_L$, producing a joint state
$\tau_{RB_1}$. By \Cref{lem:chiribella-definetti},
there is a measure-and-prepare channel whose output
$\sigma_{RB_1}$ is separable across $R:B_1$ and satisfies
\[
 \frac12\|\tau_{RB_1}-\sigma_{RB_1}\|_1
 \le \frac{d}{L+d}.
\]

Let $\tau_{\mathrm{all}}$ be the witness retained for $\mathcal P$ and $\sigma_{\mathrm{all}}$ the comparison state
obtained using the measure-and-prepare channels. Applying the single-block bound to all $N(k-1)$
replacements and using contractivity of trace distance and the triangle inequality, we obtain
\[
 \frac12\|\tau_{\mathrm{all}}-\sigma_{\mathrm{all}}\|_1
 \le N(k-1)\frac{d}{L+d}
 \le\eta.
\]

For each of the $k-1$ decoded blocks at a node,
the state $\sigma_{B_1R}$ is separable
across $B_1:R$. 
Together with node-separability of \(\dQMAsep\), the local measure-and-prepare
channels ensure that \(\sigma_{\mathrm{all}}\) is a convex
combination of admissible product witnesses. By linearity,
\(\mathcal{P}\) accepts with probability at most \(s\).
Acceptance probabilities
differ by at most the trace distance, so acceptance
on the actual witness $\tau_{\mathrm{all}}$ is at most
$s+\eta$.

Finally, $D\le[e(L+d)/d]^d$ gives
$q=O(d\log(Nk/\eta))$. Hence
\[
 p'\le p_A+(k-1)q
    \le p_A+O\!\left(k\,2^b\log\frac{Nk}{\eta}\right).
\]
All preprocessing is local, preserving the round complexity \(r\)
and proving \eqref{eq:short-witnesses}.
\end{proof}

\clearpage

\bibliography{references}

\end{document}